\documentclass[prb,onecolumn,nobibnotes,groupedaddress]{revtex4}

\usepackage{enumerate}
\usepackage{color}
\usepackage{amssymb}
\usepackage{amsmath}
\usepackage{graphicx}
\usepackage{upgreek}
\usepackage{mathtools}
\usepackage{amssymb}
\usepackage{algpseudocode}
\usepackage{algorithm}
\usepackage{braket}

\DeclareMathOperator*{\argminA}{arg\,min}

\begin{document}

\title{An attractive way to correct for missing singles excitations in unitary coupled cluster doubles theory}
%\title{Craft singles: an economical way to correct for missing singles excitations in unitary coupled cluster doubles theory}

\thanks{This manuscript has been authored by UT-Battelle, LLC, under Contract No.~DE-AC0500OR22725 with the U.S.~Department of Energy. The United States Government retains and the publisher, by accepting the article for publication, acknowledges that the United States Government retains a non-exclusive, paid-up, irrevocable, world-wide license to publish or reproduce the published form of this manuscript, or allow others to do so, for the United States Government purposes. The Department of Energy will provide public access to these results of federally sponsored research in accordance with the DOE Public Access Plan.}
%\date{}
\author{Zachary W. Windom$^{1, 2}$, Daniel Claudino$^2$\footnote{\href{mailto:claudinodc@ornl.gov}{claudinodc@ornl.gov}}, and Rodney J. Bartlett$^1$}
\affiliation{$^1$Quantum Theory Project, University of Florida, Gainesville, FL, 32611, USA \\
$^2$Quantum Information Science Section,\ Oak\ Ridge\ National\ Laboratory,\ Oak\ Ridge,\ TN,\ 37831,\ USA}

\begin{abstract}
Coupled cluster methods based exclusively on double excitations are comparatively ``cheap" and interesting model chemistries, as they are typically able to capture the bulk of the dynamical electron correlation effects. The trade-off in such approximations is that the effect of neglected excitations, particularly single excitations, can be considerable. Using standard and electron pair-restricted $T_2$ operators to define two flavors of unitary coupled cluster doubles (UCCD) methods, we investigate the extent in which missing single excitations can be recovered from low-order corrections in many-body perturbation theory (MBPT) within the unitary coupled cluster (UCC) formalism. Our analysis includes the derivations of finite-order, UCC energy functionals which are used as a basis to define perturbative estimates of missed single excitations. This leads to the novel UCCD[4S] and UCCD[6S] methods, which consider energy corrections for missing singles excitations through fourth- and sixth-order in MBPT, respectively. We also apply the same methodology to the electron pair-restricted ansatz, but the improvements are only marginal. Our findings show that augmenting UCCD with these \textit{post hoc} perturbative corrections can lead to UCCSD-quality results. %Overall, we wish to highlight the use of different orbital choices in conjunction with perturbative energy corrections as an efficacious way to conserve quantum computing resources.
%Because the presence of infinite-order singles grants a large degree of orbital insensitivity, we stress-test our methods with restricted Hartre-Fock, MP2-optimized, Brueckener, and Kohn-Sham orbitals with two distinct functionals.
\end{abstract}

%% requires nobibnotes 

\maketitle

\subsection{Introduction}
%\cite{PhysRev.97.1509,pople1977variational,pople1987quadratic}
% Define types of electron correlation
Determination of a system's electronic structure is a fundamental requirement for the understanding and computation of properties of molecules and materials. Central to this problem is the study of electron correlation, which can be generally organized into three categories: dynamic, non-dynamic, and static. In our terminology, not assumed to be universal, static is restricted to ensuring spin symmetry conditions are satisfied, leaving non-dynamic for bond breaking. Over the years, quantum chemists have proposed many methods to capture instantaneous electron interactions that embody dynamic correlation.\cite{bartlett1978many,bartlett1981many,kucharski1986fifth} After the emergence of many-body methods like coupled cluster (CC) theory,\cite{coester1960short,paldus1978correlation,bartlett1981many,shavitt2009many,bartlett2010coupled} it became clear that such a parameterization of the wavefunction naturally excels at describing dynamic correlations. On the other hand, breaking chemical bonds is not necessarily trivial and pathological cases, such as the dissociation of N$_2$, subject to a non-correctly separating RHF reference represents a significant challenge for low-rank CC methods.\cite{laidig1987description,krogh2001general,chan2004state} Notably, CC methods are not always equipped with the machinery that is needed to recover such non-dynamic correlations. Nevertheless, it is possible to improve low-rank CC approximations, for example, by relaxing the spatial and spin  symmetry requirements of the mean-field reference, usually at the cost of sacrificing $S^2$ as a good quantum number,\cite{Margraf2017-rh} followed by restoring the wavefunction's symmetry. Alternatively, one could sequentially add higher-rank cluster operators to systematically convergence toward the full configuration interaction (FCI) limit if increasingly more expensive calculations can be afforded.

 In this context, we ask the question: can a cheap and accurate single-reference method based on the CC ansatz be designed to adequately cover the range of different electron correlation ``types''?\cite{Bartlett2007-tr} This work is particularly interested in $T_2$-based models, both for simplicity and because corrections to the correlation energy initially arise in first-order via the $T_2$ operator. Including only double excitations into the CC wavefunction leads to the CCD method, which algorithmically scales as $\mathcal{O}$($N^6$) and is the simplest flavor of CC approximation. This simplicity comes at the cost of missing single and higher-order cluster operators which can negatively impact the methods' robustness. In particular, the missed single excitation effects can largely be incorporated into $T_2$-based methods with either an orbital-optimization (OO) procedure\cite{bozkaya2011quadratically,krylov2000excited,mizukami2020orbital,sokolov2020quantum} or Brueckner orbitals\cite{sherrill1998energies,hampel1992comparison}; the former being shown to offer performance gains over standard CCSD for vibrational frequencies and bond-breaking problems.\cite{sherrill1998energies}  Such  nearly infinite-order treatment would be preferred, but even perturbative approximations for singles can have some value.\cite{raghavachari1985augmented} This idea has also been pursued in the context of the random phase approximation (RPA)\cite{ren2011beyond,ren2013renormalized,voora2019variational,joshi2024generalized} and GW methods,\cite{klimevs2015singles} after being suggested to Ren by one of the authors.\cite{rjb} It is evident from these studies that even perturbative consideration of singles excitations is important to represent correlation effects associated with the initial effects of orbital relaxation.

 % discuss variations of CCD, and their purpose. pCCD
 In bond-breaking regimes, the effects of orbital relaxation are more pronounced. If such effects are not explicit in the initial single-reference determinant - as in the case of canonical Hartree-Fock (HF) orbitals - the baseline CC method can suffer from so-called non-variational catastrophes.\cite{chan2004state} One way to circumnavigate this issue is to neglect terms in the underlying CCD equations known to exhibit erroneous behavior.  Variations of the baseline CCD method have also been proposed that are shown to be beneficial in bond-breaking regimes. The CCD0 method,\cite{bulik2015can} for example, yields better results for systems exhibiting non-dynamic and static correlation by pruning the triplet coupled excitations from the $T_2$ operator. Similarly, a variant of CCD which restricts the $T_2$ tensor to its diagonal elements - referred to as (pCCD) - has also been proposed and extensively studied.\cite{cullen1996generalized,henderson2014seniority,marie2021variational,kossoski2023seniority,kossoski2021excited,brzek2019benchmarking,boguslawski2017benchmark,henderson2015pair,nowak2021orbital} It has been argued that pCCD's success can be attributed to it being a seniority-zero method, and by its correspondence to a geminal method like general valence bond (GVB).\cite{stein2014seniority} Regardless, pCCD should be  paired with an orbital-optimization procedure to extract results that are quantitatively meaningful. Lacking that, using GVB references has proven to be reasonable.\cite{stein2014seniority,kossoski2021excited,nowak2021orbital,leszczyk2021assessing,galynska2024benchmarking,Ravi2023}

 %Discuss variations of CCD from alternative ans\"{a}tze CC
As opposed to omitting terms in the $T_2$ method, one could conceive of an alternative philosophy wherein the standard $T_2$ operator is augmented with additional, higher-order terms, e.g., arising from $m$ products of $T_2$, which could feasibly explore sectors of Hilbert space associated with $2m$-fold electron excitations. This is the strategy pursued by recent work on an ``ultimate", expectation-value-based CCD method, which perturbatively expands the energy functional $\braket{0|e^{T_2^{\dagger}}H_Ne^{T_2}|0}_c$. It was found that such models are in better agreement with FCI for significant non-dynamic correlation.\cite{ultT2inc,strongCorrT2} Using this type of reasoning, an alternative strategy would be to constrain the parameterization of the CC wave-operator to be unitary leading to the UCCD ans\"{a}tze of the form $e^{\tau_2}\ket{0}$, $\tau_2 = T_2-T_2^{\dagger}$. Based on the suite of alternative ans\"{a}tze in CC theory proposed and extensively studied by Bartlett and co-workers,\cite{kaldor1989many,bartlett1989alternative,szalay1992alternative,szalay1995alternative,piecuch1999eomxcc}  such methods have historically been made tractable only by truncating the resulting infinite-order Hamiltonian-cluster operator expansion by invoking many-body perturbation theory arguments,\cite{kutzelnigg1991error,watts1989unitary,bartlett1989some} with notable alternatives.\cite{liu2022quadratic,taube2006new} Among other issues, this typically results in the method losing its variational upper bound. However, the UCC paradigm in particular is being revisited on quantum hardware which considers a unitary ans\"{a}tze implicitly at infinite-order. 
%However, with the prospects afforded by quantum computing, such paradigms can be revisited as this type of hardware . 

In fact, UCC methods are of significant interest for use on noisy intermediate scale quantum (NISQ)  computers,\cite{romero2018strategies, Claudino2022, anand2022quantum} for many of the same reasons that motivate the use of variations of standard CC on a classical computer. Notably, it is expected that electronic structure methods may only be able to take advantage of NISQ computers by constraining the excitation rank, thereby reducing the number of operations to be performed and the associated depth in the state preparation circuit. Consequently, any variation of UCC which strives to minimize the maximum rank of the cluster operator is of immediate interest. With this in mind, UCCD and electron-pair restricted UCCD (pUCCD) are particularly attractive choices for use on NISQ hardware. However, both UCCD and pUCCD ignore correlation effects due to singles excitations, a shortcoming which is typically remedied with alternative reference choices like those provided in an orbital optimization procedure.\cite{mizukami2020orbital,sokolov2020quantum,Zhao2023}

%Further improvements over baseline p/UCCD can materialized if generalized $T_2$ operators\cite{nakatsuji1976equation,nooijen2000can,nakatsuji2000structure,davidson2003exactness,mukherjee2004some} (operators that move electrons/holes within the occupied-occupied and virtual-virtual blocks) are included.\cite{lee2018generalized,kohn2022capabilities}
%Unfortunately, either route to recover missed correlation in p/UCCD means that significantly more quantum resources are required to perform the calculation, and is a bottleneck on NISQ hardware. 

With these considerations in mind, we rephrase our earlier inquiry within the context of UCC: can a cheap, single-reference unitary coupled cluster method be designed that is sufficiently accurate for the various ``types" of electron correlation? Given the recent interest in UCC due to its prospects in quantum computing, the current work addresses this issue from the perspective of providing a formally sound framework while conserving (limited) quantum hardware resources, in an effort to systematically improve the quality of UCCD results. Once in possession of converged, infinite-order $\tau_2$ amplitudes, which could in principle be obtained from a quantum computer, one is thus in position to construct perturbative energy corrections to account for the effect of missing singles excitations on a classical computer. Inspired by the long-standing success of non-iterative corrections afforded by MBPT, in what follows we introduce two singles corrections under the aptly named acronyms UCCD[4S] and UCCD[6S].% In this way, the calculation burden is shared between quantum and classical computers. 

\subsection{Theory}
Assigning the indices $a,b,c, d\cdots$ to virtual, $i,j, k, l\cdots$  to occupied, and $p,q,r,s\cdots$ to arbitrary (occupied or virtual) spin-orbitals, the normal-ordered Hamiltonian can be defined as
\begin{equation}
\begin{split}
\label{eq:ham}
&H_N=\underbrace{\sum_{p}\epsilon_{pp}\{p^{\dagger}p\}}_{\text{$f_N$}}\\
  &+ \underbrace{\sum_{ia}f_{ia}\{i^{\dagger}a\}+\sum_{ai}f_{ai}\{a^{\dagger}i\} +\frac{1}{4}\sum_{pqrs}\braket{pq||rs}\{p^{\dagger}q^{\dagger}sr\}}_{\text{$W_N$}},
  \end{split}  
\end{equation}
 where we note that  the perturbation $W_N$ now contains the occupied/virtual blocks of the Fock operator $f$.

%where $p^\dag$ and $p$ represent creation and annihilation operators, respectively, and $f_N$ and $W_N$ are the Fock operator and the two-electron part of the Hamiltonian.

UCC theory parameterizes the wavefunction using an antisymmetric wave operator, which acts on a mean-field reference $\ket{0}$ such that
\begin{equation}
    \ket{\Psi_\text{UCC}} = e^{T-T^{\dagger}}\ket{0},
\end{equation} 
where $T$/$T^{\dagger}$ are formally composed of all single (S), double (D), etc excitation/deexcitation cluster operators. The current work focuses exclusively on doubles approximations to UCC such that $T=T_2$.

In an effort to improve the tractability of UCC theory, we have recently proposed \textit{post hoc} energy corrections based on finite-order energy functionals that are guaranteed to be correct through some order in perturbation theory.\cite{windom2024new} Following the relevant fourth-order working equations, we see that the corresponding energy functional is
\begin{equation}\label{eq:finalUCC4eqn}
   \Delta E(4) = \braket{0|W_NT_2|0} +\braket{0|W_NT_1|0} - \frac{1}{4}\bigg(\braket{0|(T_1^{\dagger})^2W_NT_1|0} + \braket{0|(T_2^{\dagger})^2W_NT_2|0}\bigg),
\end{equation}
and the set of residual equations are expressed as 
\begin{subequations}\label{eq:UCC4residEqns}
        \begin{align}
     \frac{ \partial\Delta E(4)}{\partial T_1^\dagger} = 0 \Rightarrow &D_1T_1=W_N + W_NT_2 + W_NT_1 + \frac{1}{2}\big(\frac{1}{2}W_NT_1^2+T_1^{\dagger}W_NT_1\big)  +T_2^{\dagger}W_NT_2,\\
     \frac{ \partial\Delta E(4)}{\partial T_2^\dagger} = 0 \Rightarrow &D_2T_2=W_N + W_NT_2 + W_NT_1 + \frac{1}{2}\big(\frac{1}{2}W_NT_2^2+T_2^{\dagger}W_NT_2\big) +W_NT_3+T_1^{\dagger}W_NT_2+W_NT_1T_2, \\
     \frac{ \partial\Delta E(4)}{\partial T_3^\dagger} = 0 \Rightarrow &D_3T_3=W_NT_2.
    \end{align}
\end{subequations}
The complete derivation can be found in Ref. \citenum{windom2024new}. Although that work focuses on triple excitation corrections, a similar philosophy can be employed here  for the purpose of improving the energy of UCCD by compensating for missed single excitations. This procedure is prudent, since any flavor of UCCD completely bypasses solution of the $T_1$ residual equations like those found in Equation \ref{eq:UCC4residEqns}a.

\begin{figure}[ht!]
\centering
\includegraphics[width=\columnwidth]{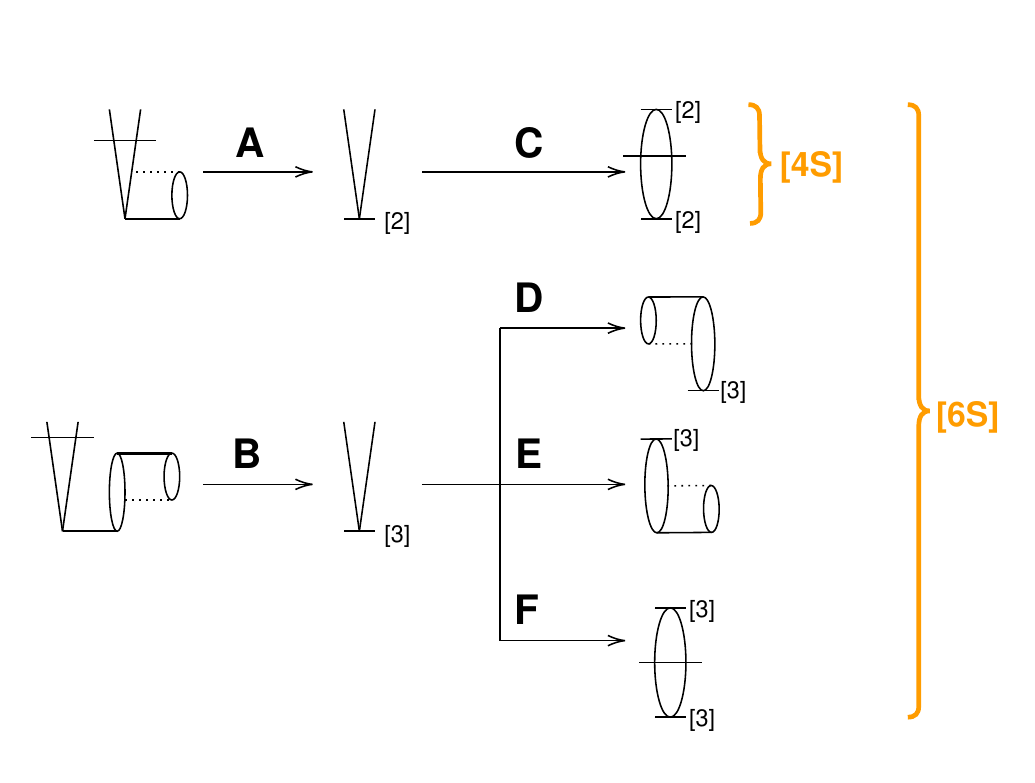}    

\caption{Outline of the procedure to extract singles' energy corrections from infinite-order $\tau_2$ amplitudes. An approximation to $T_1$, correct through second-order in MBPT is found by A) approximating $T_1$ with $\frac{(W_NT_2)_C}{D_1}$ and then C) using this quantity and the corresponding residual expression to define the fourth-order [4S] energy correction. A third-order approximation is found by B) approximating $T_1$ by $\frac{(T_2^{\dagger}W_NT_2)_C}{D_1}$. This immediately leads to higher-order energy expressions by taking all the possible contractions between $T_1^{[3]}$ and $T_1^{[2]}$ (with their corresponding residual equations) shown in diagrams D and E. An additional possibility is available based  exclusively on contracting $T_1^{[3]}$ with its respective residual equation, and is shown in diagram F. The combination of diagrams D-F define the [6S] energy correction.}
\label{fig:squarebrackS}
\end{figure}

A perturbative strategy to compensate for this deficiency can be designed as follows. Upon convergence of the infinite-order UCCD equations, we use the corresponding amplitudes to solve the $T_1$ equation found in Equation \ref{eq:UCC4residEqns}a
\begin{equation}\label{eq:T1resid}
    T_1^{[2]}=\frac{(W_NT_2)_C}{D_1}
\end{equation} and illustrated in Figure \ref{fig:squarebrackS}A. The superscript found in Equation \ref{eq:T1resid} denotes a $T_1$ that is correct through second-order in MBPT. The approximated expression for $T_1$ can then be inserted into the energy functional of Equation \ref{eq:finalUCC4eqn} such that
\begin{equation}
    \begin{split}
        \Delta E(T_1^{[2]}) &= \braket{0|f_{ia}T_1^{[2]}|0} \\
        & \approx \braket{0|T_1^{[2]\dagger}D_1T_1^{[2]}|0} = \braket{0|T_1^{[2]\dagger}(W_NT_2)_C|0},
    \end{split}
\end{equation}
where, following the convention of Ref. \citenum{windom2024new}, the last line is used to  define the [4S] perturbative energy correction; this is illustrated diagrammatically in Figure \ref{fig:squarebrackS}B. Note that the [4S] correction represents a fourth-order energy correction in MBPT.

Three, higher-order terms can be extracted by following the $(T_2^{\dagger}W_NT_2)_C$ term in Equation \ref{eq:UCC4residEqns}a. We initially see that 
\begin{equation}
    T_1^{[3]}=\frac{(T_2^{\dagger}W_NT_2)_C}{D_1},
\end{equation} which is shown diagrammatically in Figure \ref{eq:UCC4residEqns}C. Since we are looking specifically for corrections of the form $\braket{0|T_1^{\dagger}D_1T_1|0}$, there are three expressions that can be generated by interchanging $T_1$ with $T_1^{[2]}$ or $T_1^{[3]}$. These are shown to be 
\begin{equation}
    \underbrace{\braket{0|T_1^{[2]\dagger}D_1T_1^{[3]}|0} + \braket{0|T_1^{[3]\dagger}D_1T_1^{[2]}|0}}_{\Delta E^{[5]}} + \underbrace{\braket{0|T_1^{[3]}D_1T_1^{[3]}|0}}_{\Delta E^{[6]}}
\end{equation} where we highlight the fifth- and sixth-order energy corrections associated with missing single excitations. These are further illustrated sequentially in Figure \ref{eq:UCC4residEqns}D-F. The combination of diagrams B, D, E, and F in Figure \ref{eq:UCC4residEqns} define the [6S] energy correction.

%\subsection{Computational Details}

\subsection{Results}

In the following discussion, all pUCCD/UCCD calculations are performed in the XACC software,\cite{mccaskey2020xacc, xacc_chem} using a numerical simulator and relying on PySCF\cite{sun2018pyscf} for the required molecular integrals. The STO-6G basis set is used throughout this work,\cite{Hehre1969,Feller1996,Pritchard2019} and all examples drop core electrons from the correlation calculation. Once $\tau_2$ are obtained, the subsequent perturbative corrections are generated using the UT2 software.\cite{UT2}

 Motivated by the efforts in alleviating the computational burden on current noisy quantum hardware,  we turn to two flavors of UCC ans\"{a}tze solely comprised of double excitations, namely UCCD and its electron-pair restricted variant, pUCCD. Furthermore, we are interested in two main forms of the UCC ansatz. The first follows closely from the standard CC formalism where we simply replace the routine $T_2$ cluster operator by its anti-Hermitian analog $\tau_2$ and construct the exponential wave operator:

 \iffalse
 which have been optimized on a quantum simulator:
\begin{itemize}
    \item Standard UCCD
    \item Electron-pair restrictions to UCCD, leading to pUCCD
\end{itemize}
In an effort to reduce the resource burden on NISQ hardware and improve UCCD/pUCCD accuracy, we leverage a hybrid computing workflow wherein a quantum simulator converged UCC amplitudes which are subsequently leveraged by a classical computer to construct perturbative energy corrections which intend to compensate for missed electron correlations that result from the restriction $T\equiv T_2$. 
\fi

%  The error is quantified by using the formula $E_{UCC method}-E_{FCI}$, where $E_{UCC method}$ is the pertinent UCC value and $E_{FCI}$ is the exact result. All results are recorded in a spreadsheet attached in the supplementary material. Molecular geometries, at stretched and equilibrium geometries, are also included in the supplementary material. 

\begin{equation}
    \label{eq:uccd}
    |\Psi_\text{UCCD}\rangle = e^{ \sum_{ijab}\theta_{ij}^{ab}(a^\dagger b^\dagger ij- \text{h.c.})} |0\rangle.
\end{equation}
The other is based on the Trotterized (t) form of the above:
\begin{equation}
    %\prod_{k}e^{\theta_k \big(T_{1k}-T_{1k}^{\dagger}\big)}\prod_{k}e^{\theta_k \big(T_{2k}-T_{2k}^{\dagger}\big)}
     |\Psi_\text{tUCCD}\rangle = \prod_{\substack{I<J \\ A<B}}e^{\theta_{IJ}^{AB}(A^\dagger B^\dagger IJ- \text{h.c.})}e^{\theta_{\bar{I}\bar{J}}^{\bar{A}\bar{B}}(\bar{A}^\dagger \bar{B}^\dagger \bar{I}\bar{J}- \text{h.c.})} \prod_{IJAB}e^{\theta_{I\bar{J}}^{A\bar{B}}(A^\dagger \bar{B}^\dagger I\bar{J} - \text{h.c.})} |0\rangle,
    \label{eq:ansatz}
\end{equation}
with $I, J, A, B$ indexing $\alpha$ orbitals and $\bar{I}, \bar{J}, \bar{A}, \bar{B}$ indexing the corresponding $\beta$ orbitals.

The variational quantum eigensolver (VQE)\cite{Peruzzo2014} is used to obtain the $\tau_2$ amplitudes that minimize the expectation value of the Hamiltonian in Equation \ref{eq:ham}
\begin{equation}
     \tau_2^* =  \argminA_{ \tau_2}\braket{\Psi( \tau_2)| H |  \Psi( \tau_2)},
\end{equation}
with the final, converged $\tau_2^*$ being used to construct the pertinent perturbative energy corrections illustrated in Figure \ref{fig:squarebrackS}.

\subsubsection{Single-point energies}
\label{ssec:single-point}

\begin{figure}[ht!]
\centering
\includegraphics[width=\columnwidth]{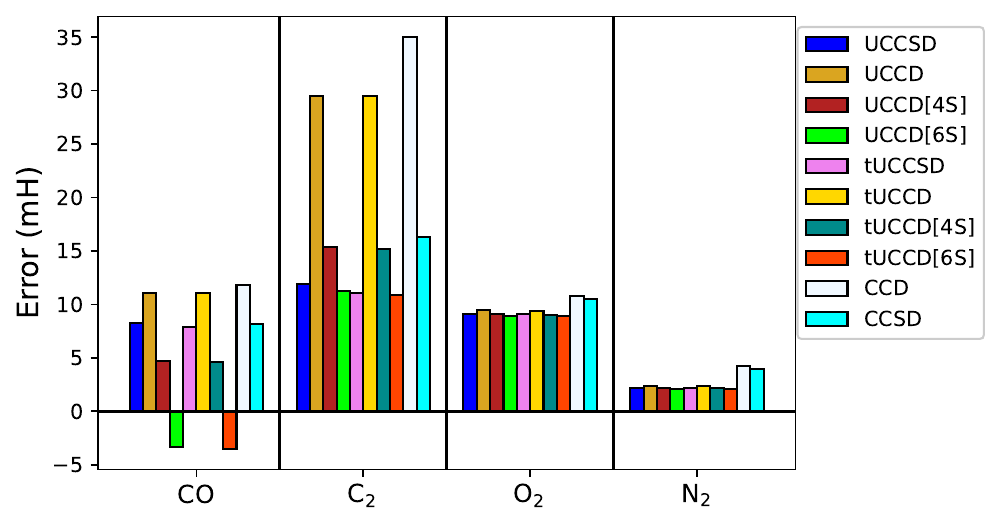}    

\caption{Errors from FCI for molecules in their experimental equilibrium geometry.}
\label{fig:EquilBarPlot}
\end{figure}

\begin{figure}[ht!]
\centering
\includegraphics[width=\columnwidth]{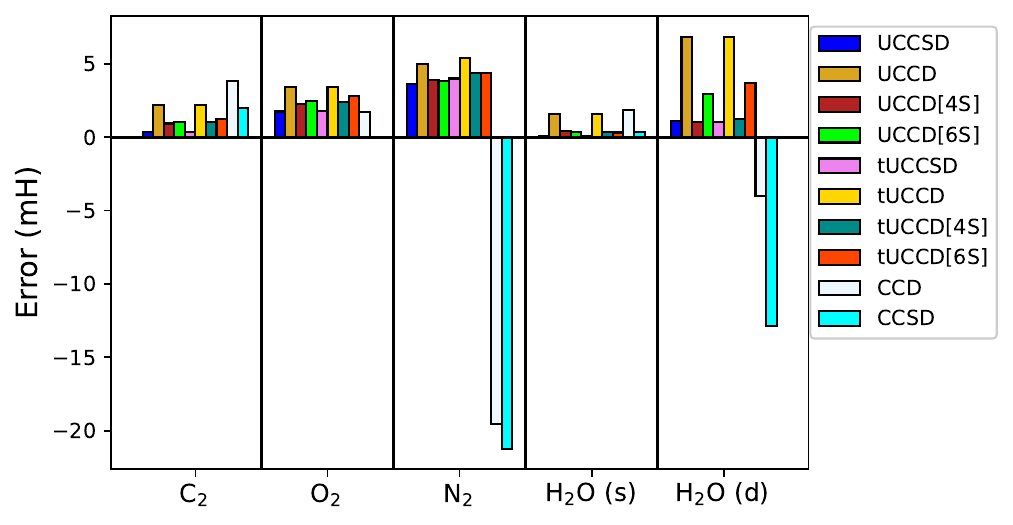}    

\caption{Errors from FCI for molecules with stretched bonds with the following lengths: $r_\text{C-C}$ = 2.243{\AA}; $r_\text{N-N}$ = 1.8{\AA}; $r_\text{O-O}$ = 1.8{\AA}; $r_\text{O-H}$ = 1.958{\AA}.The two different configurations of the H$_2$O molecule refer to a single (s) and a double (d) stretching.}
\label{fig:StretchedBarPlot}
\end{figure}

We start our analysis by looking into the inclusion of single excitation effects in a few small molecules at their equilibrium geometries as well as with their bonds stretched. A cursory inspection of Figures \ref{fig:EquilBarPlot} and \ref{fig:StretchedBarPlot} makes a few general trends apparent. Overall, tUCCD and UCCD results are reliably better than standard CCD. Similarly, it is evident that tUCCSD and UCCSD are superior to standard CCSD regardless of geometry (with the only exception being stretched O$_2$). Except in the cases of double stretched H$_2$O and stretched N$_2$, it is clear that CCSD results in a marked improvements over CCD. At least for stretched N$_2$, this result is undoubtedly an artifact of RHF-based CCSD diverging more rapidly than CCD. On the other hand, tUCCSD/UCCSD represents a broad improvement over tUCCD/UCCD. This indicates the importance of singles excitations in the underlying CC/UCC method.

We start by looking into CO, C$_2$, O$_2$, and N$_2$ in their equilibrium geometries and plot the deviation in the energy with respect to FCI in Figure \ref{fig:EquilBarPlot}. Two main conclusions can be drawn from these results. The first is that both [4S] and [6S] corrections display remarkable performance with a great extent of error cancellation, to the point that they are noticeably preferable to the inclusion of infinite-order singles, are far outperform ans\"{a}tze with single excitations alone. While [6S] is a sixth-order correction, and thus intuitively superior to [4S], its gains seem only marginal. We point out that there is no appreciable change in the energies obtained with either ansatz (Equations \ref{eq:uccd} or \ref{eq:ansatz}), which is in line with the findings reported in Ref. \citenum{windom2024new}.

Now considering molecules away from the equilibrium region, we see that as is a stress test for non-variational methods, as CCSD and CCD not only overshoot the FCI results but also dictate the scale of the plots in Figure \ref{fig:StretchedBarPlot}. Fortuitously, despite our [4S] and [6S] corrections being sizable, they do not violate the variational bound, regardless of the system. While at equilibrium one sees little difference between [4S] and [6S], this trend surprisingly holds for stretched geometries, at least for the cases that were investigated, with the only outlier to this trend being double-stretched H$_2$O.

In general, from Figures \ref{fig:EquilBarPlot} and \ref{fig:StretchedBarPlot} one is led to believe that adding either the [4S] or [6S] correction to tUCCD/UCCD results in an unanimous improvement over the baseline method. Indeed, the [4S]/[6S] corrections seem to recover most of the missing singles contribution as evident by the similarity between tUCCSD/UCCSD and [4S]/[6S]-corrected tUCCD/UCCD. This is quantified in Tables \ref{table:EquilTable} and \ref{table:StretchTable}, which record the  total percentage of correlation extracted by each method. Here, we see that for equilibrium CO and C$_2$, adding either singles correction to tUCCD/UCCD provides $>$4\% improvement in the results over the corresponding $T_2$-only method; and in either case directly competes with tUCCSD/UCCSD. Although the difference in correlation contribution between tUCCD/UCCD and tUCCSD/UCCSD are small in the remaining equilibrium examples, the [4S]/[6S] correction to tUCCD/UCCD nevertheless results in tUCCSD/UCCSD-quality results.

Curiously, the quality of both tUCCSD/UCCSD and tUCCD/UCCD generally improves for stretched geometry examples, providing $>$98\% of the total correlation contribution except in the case of stretched CO. This is somewhat of an anomaly, as the Trotterized methods (Equation \ref{eq:ansatz}) actually provide better quality results than the full operator analogs (Equation \ref{eq:uccd}).  Whereas the differences between [4S] and [6S] are generally within $<1$\% of each other, the differences are more pronounced in the case of CO. Adding the fourth-order [4S] correction to tUCCD/UCCD recovers roughly 20\% extra correlation. The sixth-order [6S] correction for this example does even better, recovering at least 30\% extra correlation beyond the tUCCD/UCCD method.

Together with the results discussed so far, Tables \ref{table:EquilTable} and \ref{table:StretchTable} also report results with the electron-pair restricted tUCCD/UCCD, referred to as tpUCCD/pUCCD, which are overall less appealing. Baseline tpUCCD/pUCCD using canonical HF orbitals routinely underestimates the FCI by a large margin. As previously mentioned, it is now common knowledge that this flavor of methods needs to be augmented with alternative reference choices - like Brueckner or optimized orbitals\cite{sherrill1998energies,scuseria1987optimization,krylov2000excited} - to yield quantitatively meaningful results. Such techniques would naturally incorporate elements of singles excitations into tpUCCD and pUCCD. This begs the question of how close to FCI we can get with singles energy corrections added onto tpUCCD or pUCCD. Although the [4S]/[6S]-corrected tpUCCD/pUCCD methods show consistent improvement as opposed to the absence of any single corrections, the results are not appreciable enough to routinely compete with tUCCSD/UCCSD or even tUCCD/UCCD. This is somewhat intuitive, as the single correction for tpUCCD/pUCCD is constructed from a $T_2$ tensor which only has diagonal elements, placing severe restrictions on the number of low-order single amplitudes that can be constructed, and subsequently the amount of extra singles' correlation that can be recovered. 

\begin{table}[ht!]
\begin{center}
\caption{Percentage of correlation energy recovery from the methods considered in this work. The molecules are at their equilibrium geometries (Ref. \citenum{cccdbd}).}
\begin{tabular}{cccccc} 
 \hline
	&C$_2$	&CO	&H$_2$O&	N$_2$&	O$_2$\\ \hline \hline
UCCSD	&95.62	&94.08	&99.80	&98.63	&94.15\\
UCCD	&89.14	&92.08	&99.32	&98.49	&93.96\\
UCCD[4S]	&94.34	&96.63	&99.69	&98.64	&94.19\\
UCCD[6S]	&95.85	&102.4&	99.75&	98.69&	94.28\\
pUCCD	&58.65	&46.77	&50.58	&43.53	&75.56\\
pUCCD[4S]&	60.65	&48.09&	50.70&	43.56&	75.66\\
pUCCD[6S]&	60.66	&48.70&	50.70	&43.55&	75.61\\
tUCCSD	&95.92	&94.34	&99.80	&98.63	&94.16\\
tUCCD	&89.14	&92.08	&99.32	&98.49	&93.97\\
tUCCD[4S]&	94.40	&96.67	&99.69&	98.64&	94.21\\
tUCCD[6S]&	95.98	&102.5	&99.75&	98.70&	94.32\\
tpUCCD	&53.70	&46.17	&50.58	&50.31	&75.56\\
tpUCCD[4S]&	55.68&	47.45	&50.70&	50.34&	75.66\\
tpUCCD[6S]&	55.69&	48.03	&50.7	&50.32	&75.61\\
CCD	&87.10	&91.53&	99.26	&97.33	&93.07\\
CCSD&	93.98&	94.10&	99.76&	97.49	&93.30\\ \hline
\end{tabular}
\label{table:EquilTable}
\end{center}
\end{table}

\begin{table}[ht!]
\begin{center}
\caption{Percentage of correlation energy recovery from the methods considered in this work. The molecules are at the geometries presented in the caption of Figure \ref{fig:StretchedBarPlot}. The two different configurations of the H$_2$O molecule refer to a single (s) and a double (d) stretching.}
\begin{tabular}{ccccccc} 
 \hline
	&C$_2$	&CO&	N$_2$&	O$_2$ &	H$_2$O(s)& 	H$_2$O(d)\\ 
 \hline \hline
UCCSD	&	99.89	&	74.01	&	99.23	&	99.55	&	99.95	&	99.68	\\
UCCD	&	99.31	&	50.76	&	98.93	&	99.12	&	99.18	&	98.03	\\
UCCD[4S]	&	99.71	&	69.52	&	99.16	&	99.41	&	99.78	&	99.69	\\
UCCD[6S]	&	99.68	&	82.32	&	99.18	&	99.36	&	99.81	&	99.15	\\
pUCCD	&	81.79	&	23.81	&	82.02	&	91.69	&	92.95	&	57.80	\\
pUCCD[4S]	&	82.51	&	24.49	&	82.32	&	92.18	&	94.48	&	58.58	\\
pUCCD[6S]	&	82.34	&	24.28	&	82.16	&	92.05	&	94.53	&	58.17	\\
tUCCSD	&	99.90	&	86.55	&	99.15	&	99.55	&	99.96	&	99.71	\\
tUCCD	&	99.31	&	50.78	&	98.85	&	99.12	&	99.18	&	98.03	\\
tUCCD[4S]	&	99.68	&	70.43	&	99.07	&	99.38	&	99.80	&	99.65	\\
tUCCD[6S]	&	99.61	&	84.20	&	99.07	&	99.28	&	99.84	&	98.94	\\
tpUCCD	&	81.79	&	22.93	&	81.48	&	91.69	&	92.95	&	57.79	\\
tpUCCD[4S]	&	82.51	&	23.45	&	81.78	&	92.18	&	94.48	&	58.56	\\
tpUCCD[6S]	&	82.34	&	23.31	&	81.62	&	92.03	&	94.53	&	58.16	\\
CCD	&	98.79	&	49.94	&	104.2	&	99.56	&	99.05	&	101.2	\\
CCSD	&	99.36	&	81.08	&	104.5	&	100.0	&	99.81	&	103.7	\\
 \hline
\end{tabular}
\label{table:StretchTable}
\end{center}
\end{table}

\subsection{Potential energy curves}

Breaking chemical bonds inherently involves correlations of different ``types", particularly when the single-reference determinant no longer has sizable overlap with the exact wavefunction. As higher excitation manifolds are included in the CC ansatz, it eventually blurs the line dividing these disparate types of correlations until we systematically recover the FCI limit. With this in mind, in comparing the results directly to FCI we are unable to disregard higher-excitations effects, which can be sizable. At the same time, even though intuitively one may assume the inclusion of infinite-order singles should be the target, the MBPT series is not monotonic and the trend in error cancellation may not be easily predictable. This is in line with the conclusions drawn in Subsection \ref{ssec:single-point}. To better understand the strengths and weaknesses of our proposed singles' corrections, we study the potential energy curves (PEC) of LiF and linear H$_8$. These systems were chosen as representatives of different ``types'' of correlation, since we are breaking a single and resonant quadruple bond, respectively. For these examples, it continues to be the case that tUCCSD/UCCSD results represent improvements over baseline tUCCD/UCCD. 

\begin{figure}[ht!]
\centering
\includegraphics[width=\columnwidth]{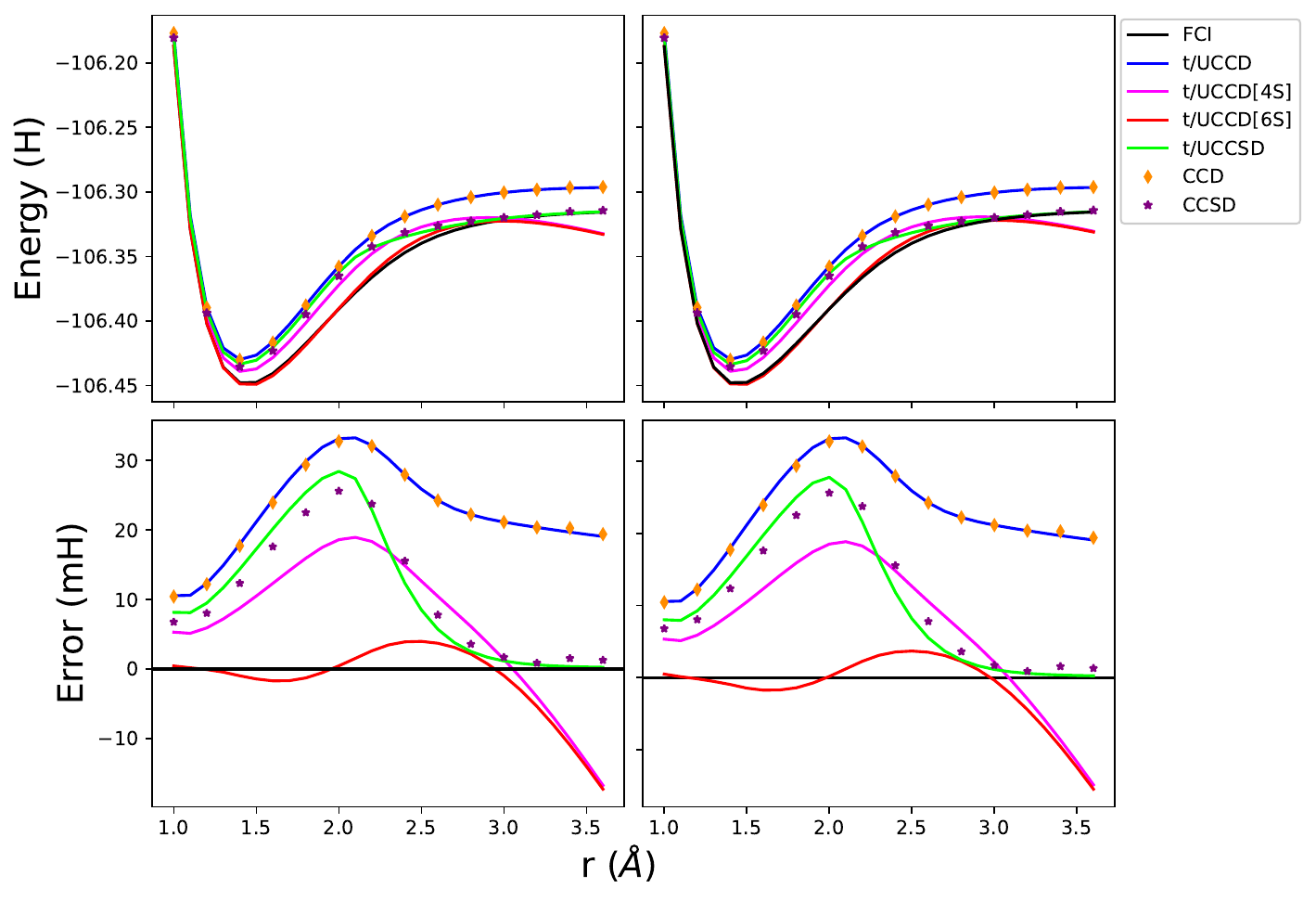}    

\caption{A comparison of UCC methods for the dissociation of LiF with the UCCD ansatz (left) and Trotterized UCCD ansatz (right).}
\label{fig:LiF}
\end{figure}

The dissociation of LiF is shown in Figure \ref{fig:LiF}, where it is clear that adding singles' energy corrections to tUCCD/UCCD offer tangible benefits. Judging from the magnitude of the deviation from FCI alone, we see that both [4S] and [6S] corrections outperform the inclusion of infinite-order singles up to  $\approx$2.2 {\AA}. Additionally, in this range of distances, while [6S] shows smaller errors, [4S] tends to keep the total energy above FCI for longer bond lengths, which is also observed for the other two molecules discussed later on. Beyond this point, UCCSD and its Trotterized variant overcome [4S], with this only happening for [6S] in the vicinity of 3{\AA}.  That is, in the range of $\approx$2.2-3.0{\AA} it turns out the tUCCD[6S]/UCCD[6S] has the best performance. Beyond 3.0{\AA} both of [4S] and [6S] energy corrections to tUCCD/UCCD lead to underestimation of the FCI energy, a signature of perturbative corrections. 

%Figure \ref{fig:CO} depicts the PEC of CO, where it is clear that the error of both tUCCD/UCCD and tUCCSD/UCCSD increases with bond length, although both are upper bounds to the FCI as expected. As in the case of LiF, it also appears to be the case for tUCCD[S]/UCCD[S], although this is a fortuitous happenstance since there is no formal guarantee that such perturbative corrections obey a variational principle. Practically, this behavior can be seen for the majority of the PEC. On the other hand, tUCCD(S)/UCCD(S) largely underestimates FCI but is nevertheless a clear improvement over baseline tUCCD/UCCD.

The symmetric dissociation of linear H$_8$ is a pathological example where traditional, projective CC theory dramatically fails. In such cases of ``strong" correlation, the variational properties of UCC are expected to pay dividends. We find marginal improvement for this example in moving from tUCCD/UCCD to tUCCSD/UCCSD. Again, we note that over the range of the PEC both [4S] and [6S] corrections to tUCCD/UCCD show better agreement with FCI at least until 2.5 {\AA}. At this point, both methods begin to underestimate FCI, at the largest bond distance - 3.0 \AA - the overall errors between UCCSD and UCCD[4S] are comparable. In this case, it is curious that the Trotterized methods are in overall better agreement with FCI than the standard CC ansatz alternatives.

\begin{figure}[ht!]
\centering
\includegraphics[width=\columnwidth]{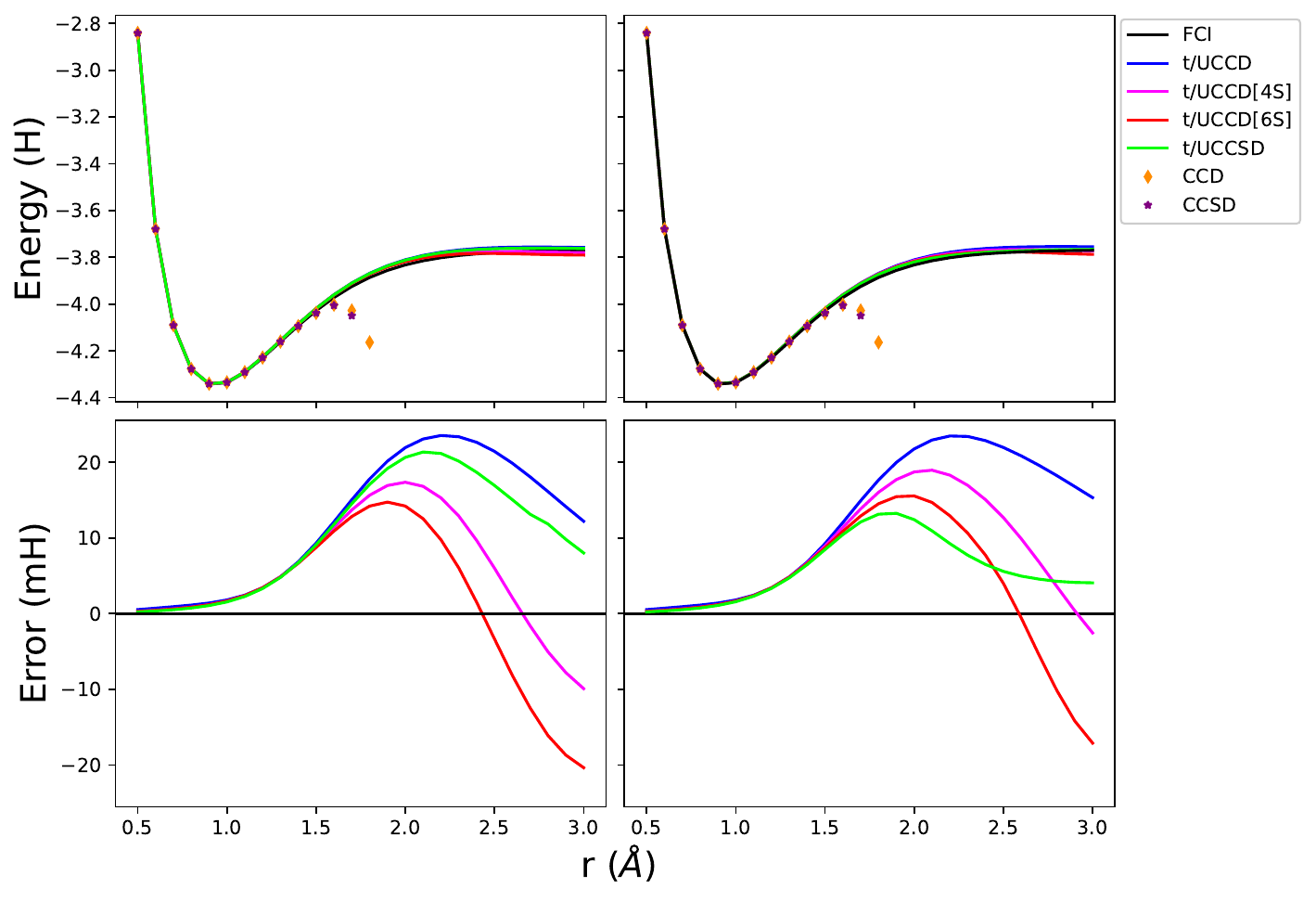}    

\caption{A comparison of UCC methods for the symmetric dissociation of H$_8$ with the UCCD ansatz (left) and Trotterized UCCD ansatz (right).}
\label{fig:H8}
\end{figure}

\subsection{Conclusion}

Electronic structure methods that can efficiently describe a multitude of electron correlation ``types" represent a ``holy grail" of quantum chemistry. For situations where the correlations are considered strong, methods based on single-reference coupled cluster theory require inclusion of higher-rank cluster operators to compensate which has the unfortunate side-effect of significantly increasing calculation cost. Thus, a balance between method robustness and expense must be struck. Although single-reference UCC theory is supported by a variational bound to the exact answer, which endows it with benefits not found in traditional CC theory, UCC calculations continue to be expensive to perform at scale on either classical or quantum computers. To this end, the current work seeks to retroactively correct the cheapest UCC method - UCCD - for missed singles excitations. Following the tone set in our previous work (Ref. \citenum{windom2024new}) and envisioning its usage in the context of small and noisy quantum hardware, our proposed solution points to yet another independent and complementary way to manage calculation expense by appropriately leveraging quantum and classic hardware.

The current analysis seems to indicate that there is little downside to adding perturbative corrections to either tUCCD or UCCD. In fact, it is shown  in several of the examples provided that perturbative singles corrections to tUCCD/UCCD are actually closer to FCI than the infinite-order tUCCSD/UCCSD methods. Similarly, perturbatively correcting for singles using amplitudes from the electron-pair restricted ans\"{a}tze consistently improve the underlying results. However, the resulting correction is mostly quantitatively inconsequential in the energy scale of the observed error, and the corresponding singles corrected tpUCCD/pUCCD methods still remain far from the FCI energy.

% discuss limited basis set size

As with all perturbative techniques, it is not always clear when such methods will meet certain expectations in quantum chemistry nor is it always apparent which particular flavor of correction will lead to the best answer for a particular problem. Although traditional, low-rank CC techniques using MBPT-based energy corrections to accommodate higher-rank operators will struggle in pathological situations, like the symmetric dissociation of linear H$_8$, we find that the energy corrections using tUCCD/UCCD amplitudes are far more robust than our intuition might otherwise dictate, and the current results are encouraging for ``real'' molecules. Future investigations considering the extent to which these results follow from the use of a minimal basis set are clearly warranted. While such \textit{post hoc} corrections forgo a variational upper bound to the FCI, this work suggests that this might be a worthwhile sacrifice to extract better agreement - at least for a portion of the PEC. At extreme bond distances, these corrections begin to underestimate the FCI although with absolute errors that at least improve upon baseline tUCCD/UCCD. Here again, it would be prudent to see what effect - if any - basis set size has on these trends. Thus, future work will also focus on extending the reach of simulators with the goal of enabling calculations with larger basis sets.

% discuss role/context of quantum computing

%\clearpage

\section*{Acknowledgements}
This work was supported by the Air Force Office of Scientific Research under AFOSR Award No. FA9550-23-1-0118. Z.W.W. thanks the National Science Foundation and the Molecular Sciences Software Institute for financial support under Grant No. CHE-2136142. Z.W.W. also acknowledges support from the U.S. Department of Energy, Office of Science, Office of Workforce Development for Teachers and Scientists, Office of Science Graduate Student Research (SCGSR) program. The SCGSR program is administered by the Oak Ridge Institute for Science and Education (ORISE) for the DOE. ORISE is managed by ORAU under contract number DE-SC0014664. D.C. acknowledges support by the “Embedding Quantum Computing into Many-body Frameworks for Strongly Correlated Molecular and Materials Systems” project, which is funded by the U.S. Department of Energy (DOE), Office of Science, Office of Basic Energy Sciences, the Division of Chemical Sciences, Geosciences, and Biosciences.

%\section*{References}

\bibliography{biblio.bib}

\end{document}